\documentclass[pra,aps,floats,letterpaper,floatfix,footinbib,preprintnumbers,twocolumn]{revtex4}

\usepackage{amsmath}

\usepackage{graphicx}

\begin{document}

\title{Atomic clocks and coherent population trapping: Experiments for undergraduate laboratories}

\author{Nathan Belcher}
\author{Eugeniy E. Mikhailov}
\author{Irina Novikova}
\affiliation{Department of Physics, College of William \& Mary, Williamsburg, Virginia 23185, USA}


\begin{abstract}
We demonstrate how to construct and operate a simple and affordable apparatus for producing coherent effects in
atomic vapor and for investigating their applications in time-keeping and magnetometery. The apparatus consists
of a vertical cavity surface emitting diode laser directly current-modulated using a tunable microwave
oscillator to produce multiple optical fields needed for the observation of coherent population trapping. This
effect allows very accurate measurement of the transition frequency between two ground state hyperfine
sublevels, which can be used to construct a coherent population trapping-based atomic clock.

\end{abstract}

\maketitle

\section{Introduction}

Coherent interactions of electromagnetic fields with atoms and molecules are of much interest because they
enable coherent control and manipulation of the quantum properties of light and matter. In particular, the
simultaneous interaction of atoms with two or more light fields allows all-optical addressing of the microwave
transition between long-lived spin states of alkali metals such as Rb or Cs. If the frequency difference of the
two fields exactly matches the splitting between two hyperfine sublevels of the atomic ground state, the atoms
are prepared in a non-interacting coherent superposition of the two states, known as a ``dark state.'' This
effect is known as coherent population trapping.\cite{arimondo96PO} Because the dark state exists only for a
very narrow range of differential frequencies between the two optical fields, a narrow transmission peak is
observed when the frequency of either optical field is scanned near the resonance, and thus this effect is often
called electromagnetically induced transparency.\cite{fleischhauerRMP05} There are many important applications
of these effects such as atomic clocks,\cite{vanier05apb,comparo07pt,wynandsnature}
magnetometers,\cite{fleischhauer94PRA,hollberg04APL} slow and fast light,\cite{boydreview} quantum memory for
photons,\cite{lukin03rmp,lukin06opn} and nonlinear optics at the single photon level.\cite{lukin00prl} In recent
years electromagnetically induced transparency and related effects have gone beyond atomic systems and have been
adapted for more complex systems such as molecules, impurities in solid state crystals, quantum dots, optical
microresonators, and other photonic structures.

Although the coherent control and manipulation of atomic and light quantum properties is
becoming increasingly important in many areas of physics, there are only a few publications aimed at
introducing undergraduate physics students to the concepts.\cite{BerkeleyNMOR} The complexity and
high cost of equipment for conventional electromagnetically induced transparency are the main obstacles in making these experiments more
accessible for students with little or no experience in optics. In this paper we present an experimental
arrangement that allows undergraduate students to observe electromagnetically induced transparency and study
its properties, as well as to build an atomic clock and/or magnetometer by locking a microwave
oscillator on a clock resonance in Rb atoms.

\section{Brief summary of relevant theory}

\begin{figure}[h!]
\centering
\includegraphics[width=1.0\columnwidth]{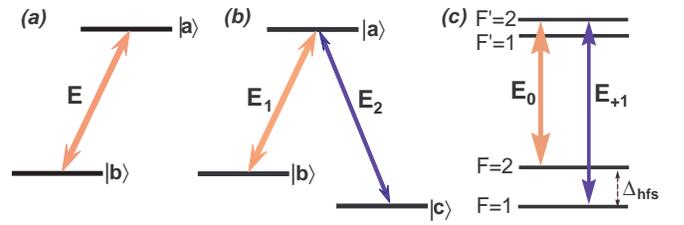}
\caption{\label{LambdaSystem.fig} Interaction of light with (a) two-level atom, (b) three-level atom in a
$\Lambda$-configuration, and (c) realization of the $\Lambda$ configuration for the $\mathrm{D}_1$ line of
${}^{87}$Rb. Here $\mathrm{E}_0$ and $\mathrm{E}_{+1}$ are the carrier and the first high-frequency modulation
sideband.}
\end{figure}

A complete analytical treatment of electromagnetically induced transparency would include spontaneous
transitions between different atomic levels and requires use of the density matrix formalism, which is not
generally introduced to undergraduates. However, the essence of the effect can be easily demonstrated using wave
functions.\cite{griffiths}

We first briefly review the important results regarding the interaction of a two-level atom with
an electromagnetic field (see Fig.~\ref{LambdaSystem.fig}(a)). In this case an
atom can be in a superposition of two atomic states $\psi_a$ and $\psi_b$, which are eigenstates of
the unperturbed atomic Hamiltonian $\hat H_0$ such that
\begin{equation}
\hat H_0 \psi_i=\hbar \omega_i \psi_i \qquad (i=a,b).
\end{equation}
Normally, all atoms are in their ground state, but an oscillating electromagnetic
field can vary the populations by exciting atoms to the an excited state. Thus, we
expect to find the atomic wave function in the form
\begin{equation} \label{psi2l}
\Psi(t)=C_a(t) e^{-i\omega_at}\psi_a + C_b(t) e^{-i\omega_bt}\psi_b,
\end{equation}
where the coefficients $C_{a,b}(t)$ are functions of time and the probability of finding an atom in this state
is $|C_{a,b}(t)|^2$. To find these coefficients we need to solve the time-dependent Schr\"{o}dinger equation
\begin{equation} \label{Schred-eqn}
i\hbar \frac{\partial \Psi(t)}{\partial t}= \hat H \Psi(t),
\end{equation}
where $\hat H = \hat
H_0 + \hat H^\prime$. The interaction part of the Hamiltonian $\hat H^\prime$ contains information about the interaction of
atoms with the electromagnetic field $E(t)=\mathcal{E}\cos(\omega t)$. When the light is nearly
resonant, that is, when its (angular) frequency $\omega$ is close to the frequency difference between two atomic states
$\omega_{ab}=\omega_a-\omega_b$, it is convenient to use the \emph{rotating wave
approximation}. This approximation neglects the fast oscillating part of the solution (proportional
to $e^{\pm i(\omega+\omega_{ab})t}$), which averages out at the detection stage, and keeps
track of only measurable slow changes on a time scale proportional to $1/(\omega-\omega_{ab})$. In
this case the only two non-zero matrix elements of the interaction Hamiltonian are\cite{griffiths}
\begin{subequations}
\begin{align}
H^{\prime}_{ab}&=\langle\psi_a|H^{\prime}|\psi_b\rangle= \hbar \Omega e^{-i \omega t},\\
\noalign{\noindent and}
H^{\prime}_{ba}&=\hbar \Omega e^{i \omega t}.
\end{align}
\end{subequations}
Here $\Omega$ is proportional to the light field amplitude:
\begin{equation} \label{Rabi}
\Omega = \frac{\wp_{ab}\mathcal{E}}{2\hbar},
\end{equation}
where the parameter $\wp_{ab} \equiv \langle\psi_a|-ez|\psi_b\rangle$ is the matrix element of the electron
dipole moment, whose value is determined by the intrinsic properties of an atom and characterizes the
strength of interaction with an external electromagnetic field.

To find the equations describing the time evolution of the state coefficients $C_a$ and $C_b$, we need to
substitute the wavefunction in Eq.~(\ref{psi2l}) into the Schr\"{o}dinger equation (\ref{Schred-eqn}) and then
use the orthogonality conditions for the wave functions, $\langle\psi_a|\psi_b\rangle = 0$, to find:
\begin{subequations}
\label{coeff2l}
\begin{align}
i \dot{C}_{a} &= \Omega e^{i(\omega-\omega_{ab})t}C_b, \\
i \dot{C}_{b} &= \Omega e^{-i(\omega-\omega_{ab})t}C_a.
\end{align}
\end{subequations}
The solution of Eq.~\eqref{coeff2l} is well-known:\cite{griffiths} the atomic population cycles between
the ground and excited states at the frequency $\sqrt{(\omega-\omega_{ab})^2+\Omega^2}$. Such
oscillations are called Rabi flopping, and the parameter $\Omega$ is usually called the Rabi frequency.

Thus far we have considered only stimulated transitions between two atomic states, which occur when the jumps
between two atomic states are caused solely by an electromagnetic field. In this case atoms
repeatedly absorb and emit photons of the incident electromagnetic field, and no energy is lost in
the process. This picture is not completely accurate, because it
does not take into account the finite lifetime of the excited state. When atoms are in the excited
states, they can spontaneously decay into the ground state by emitting a photon in a random
direction, so that the energy carried out by this photon is lost from the original light field. As a
result, some fraction of resonant light is absorbed after the interaction with atoms.

We now return to the main topic of interest for our experiment -- the interaction of three-level atoms
with two nearly-resonant laser fields, forming the $\Lambda$ configuration shown in
Fig.~\ref{LambdaSystem.fig}(b). In this case the state of such an atomic system is given
by a superposition of all three states:
\begin{equation} \label{psi3l}
\Psi(t)=C_a(t) e^{-i \omega_at}\psi_a + C_b(t) e^{-i \omega_bt}\psi_b + C_c(t) e^{-i\omega_ct}\psi_c.
\end{equation}
In the following we assume that the lifetimes of the ground states $|\psi_b\rangle$ and $|\psi_c\rangle$ are
very long, and the excited state $|\psi_a\rangle$ spontaneously decays to the ground states at the average rate
of $\gamma_a$. We also assume that each of the electromagnetic fields interacts with only one atomic transition
-- the field $E_1(t)=\mathcal{E}_{1}\cos(\omega_1 t)$ couples the states $|\psi_a\rangle$ and $|\psi_b\rangle$,
and the field $E_2(t)=\mathcal{E}_{2}\cos(\omega_2 t)$ couples the states $|\psi_a\rangle$ and $|\psi_c\rangle$.
This assumption means that the only non-zero matrix elements of the three-level system interaction Hamiltonian
are $H^\prime_{ab}=\hbar \Omega_1 e^{-i\omega_1 t}$ and $H^\prime_{ac}=\hbar \Omega_2 e^{-i\omega_2 t}$  and
their complex conjugates. If we follow the same steps as for a two-level system, we arrive at the following
equations for the coefficients of the wavefunction in Eq.~(\ref{psi3l}):
\begin{subequations}
\begin{align}
i \dot{C_a} &= \Omega_1 e^{i(\omega_1-\omega_{ab})t}C_b+\Omega_2
e^{i(\omega_2-\omega_{ac})t}C_c \label{coeff3l} \\
i \dot{C_b} &= \Omega_1 e^{-i(\omega_1-\omega_{ab})t}C_a \\
i \dot{C_c} &= \Omega_2 e^{-i(\omega_2-\omega_{ac})t}C_a.
\end{align}
\end{subequations}

We have so far neglected the finite lifetime of the excited state $|\psi_a\rangle$. To properly account for it
we have to use more sophisticated density matrix formalism.\cite{lukin03rmp} However, the optical losses due to
spontaneous emission should be proportional to the population of the excited state $|C_a|^2$. Thus, if we can
prevent atoms from getting excited, no energy will be dissipated, and the light field will propagate through
atoms without any absorption. Let's examine Eq.~(\ref{coeff3l}) more closely and find the condition for which
$\dot{C_a}(t)=0$ at all times. This condition corresponds to cases for which no atoms from either ground state
are excited at any time even in the presence of the light fields, and there is no atomic population in the
excited state ($C_a(t)=0$), and therefore no light is absorbed.  It is easy to see that this condition is
possible only when:
\begin{equation} \label{darkstate1}
\Omega_1 C_b = - \Omega_2 C_c e^{i\left[(\omega_2-\omega_{ac})-(\omega_1-\omega_{ab})\right]t}.
\end{equation}
Because the phases of laser fields are usually constant, Eq.~\eqref{darkstate1} requires that
$(\omega_2-\omega_{ac})-(\omega_1-\omega_{ab}) = 0$, which can be rewritten as
\begin{equation} \label{2photonres}
\omega_2- \omega_1 = \omega_{bc}.
\end{equation}
This condition is often called a \emph{two-photon resonance}, because it requires the difference of the
frequencies of two light fields to match the frequency splitting of two metastable states $\omega_{bc}$. For the
exact two-photon resonance the condition (\ref{darkstate1}) becomes much simpler: $\Omega_1 C_b = - \Omega_2
C_c$. If we substitute the coefficients into the general expression for the wave function (\ref{psi3l}) and
choose the proper normalization, we find that there exists a non-interacting quantum state of the atomic system
that is completely decoupled from the excited state:
\begin{equation}
|D\rangle = \frac{\Omega_2 e^{-i\omega_bt}\psi_b - \Omega_1
e^{-i\omega_ct}\psi_c}{\sqrt{\Omega_1^2+\Omega_2^2}}.
\end{equation}
Any atom in state $|D\rangle$ never gets excited to the level $|a\rangle$, and the sample
viewed from the side stays dark due to the lack of spontaneous emission. For that reason such a
quantum state is often called a ``dark state,'' in which the atomic population is ``trapped'' in two
lower states.

We can now understand what happens when an atom enters the interaction region. Before interacting
with light the atomic population is equally distributed between two low energy states; that is, half of the atoms are in the state $|b\rangle$ and half are in the state $|c\rangle$. It is
very important to distinguish this statistical mixture from the coherent superposition
of two states $(|b\rangle+|c\rangle)/\sqrt{2}$. During the interaction with light the atoms
are excited to state $|a\rangle$, and then spontaneously decay either into the dark state, or into
its orthogonal counterpart, the bright state. The atoms in the dark state do not interact with the
laser fields and remain in this state for as long as it exists. The atoms in the bright state
interact with the applied fields, so that they go through an excitation and spontaneous decay cycle many
times before all atoms end up in the dark state. After such a steady state is achieved, the atomic
medium becomes transparent, because both resonant light fields propagate without any optical
losses.
Note that the dark state is a coherent superposition of the two atomic states $|b\rangle$
and $|c\rangle$, and is sensitive to their relative phase. That exchange requires the two optical fields to
maintain their relative phase at all times in order for atoms to stay ``invisible'' to the light
fields. That is why this effect is called coherent population trapping.

We now discuss what happens to light transmission if the frequency of one of the lasers is
changing. If the system is not exactly at the two-photon resonance, a small two-photon detuning
$\delta = \omega_2- \omega_1 - \omega_{bc} \neq 0$ causes a slow variation in the relative phase of
the two ground states:
\begin{equation} \label{darkstate}
|D\rangle = \frac{\Omega_2 e^{-i\omega_bt}\psi_b - e^{i \delta\cdot t}\Omega_1
e^{-i\omega_ct}\psi_c}{\sqrt{\Omega_1^2+\Omega_2^2}}.
\end{equation}
If the two-photon detuning parameter $\delta$ is small, we can still use the dark state formalism as long as
$\delta \cdot t \lesssim 1$. At longer times such a ``disturbed dark state'' starts interacting with the laser
fields, causing non-zero atomic population in the excited state and associated optical losses due to spontaneous
emission.

So far we have allowed atoms to remain in the dark state indefinitely. In a real system some decoherence
mechanisms are present to disturb a quantum state by randomizing its phase or forcing atoms to jump between two
random energy levels. Even under perfect two-photon resonance atoms cannot be in the dark state longer than a
characteristic dark state lifetime $\tau$. Therefore, even for small non-zero two-photon detuning, the
transparency remains if the additional phase accumulated by the dark state during its lifetime is small, $\delta
\le 1/\tau$. For larger two-photon detuning the dark state no longer exists, and the amount of light absorption
becomes large. Thus we can characterize the width of the coherent population trapping resonance, that is, the
range of two-photon detunings where transmission is still high. The exact expression of the width can be found
only using a more complete treatment:\cite{vanier05apb}
\begin{equation} \label{gammawidth}
\delta_{\mathrm{CPT}}=\frac{1}{\tau}+ \frac{|\Omega_1|^2+|\Omega_2|^2}{\gamma_a}.
\end{equation}
For stronger laser fields coherent
population trapping resonance is broadened (``power broadened''), but the ultimate width is
limited by the inverse lifetime of the dark state.

We have discussed an idealized three-level atom, but no such atoms exist in
nature. It is possible to realize a three-level $\Lambda$ system in alkali metals very similar to one we have
considered. In these elements, two non-degenerate hyperfine states of the
ground $nS_{1/2}$ level are used as states $|\psi_b\rangle$ and $|\psi_c\rangle$, and the electron
state $nP_{1/2}$ or $nP_{3/2}$ becomes the excited state $|\psi_a\rangle$. Because spontaneous
radiative decay between two hyperfine states of the same ground level is strictly forbidden, an atom
can stay in either state until it interacts with its environment.

It is now possible to preserve the
quantum state of atoms for up to several seconds, but it requires using cold atoms or exotic chemical
coatings. In a regular vapor cell (a sealed glass cell filled with alkali metal vapor at about
room temperature) the main limitation of the ground state lifetime is the motion of atoms. Once an
atom leaves the laser beam, it is likely to collide with the glass wall and thermalize and lose
any information about its previous quantum state. For a $1$\,mm laser beam the interaction time of an atom
is limited to a few $\mu$s, which corresponds to a coherent
population trapping linewidth of tens or even hundreds of kHz.
Sometimes a buffer gas --- usually a non-interacting inert gas --- is added to the vapor cell together
with alkali metal. In this case alkali atoms diffuse through the laser beam rather than moving
ballistically, thus increasing the interaction time by a few orders of magnitude, producing
narrower coherent
population trapping resonances.

\section{Experimental setup}

Diode lasers are an ideal choice for working with alkali metals (K, Rb, Cs)\cite{steck} because they
are affordable, reliable, and easy to operate, and cover the right spectral range. The exact output
frequency of a diode laser can be fine-tuned by changing the driving current and/or the temperature
of the diode. In our experiments we use a laser resonant with the $\mathrm{D}_1$ line of ${}^{87}$Rb
(wavelength $\lambda=795$\,nm). A different Rb isotope or any other alkali
metal\cite{wynandsvcsel,wynandsD1D2,merimaaJOSAB03} can be used to reproduce the experiments
described in the following with an appropriate change in the operational parameters.

\begin{figure*}[h!]
\includegraphics[width=0.9\textwidth]{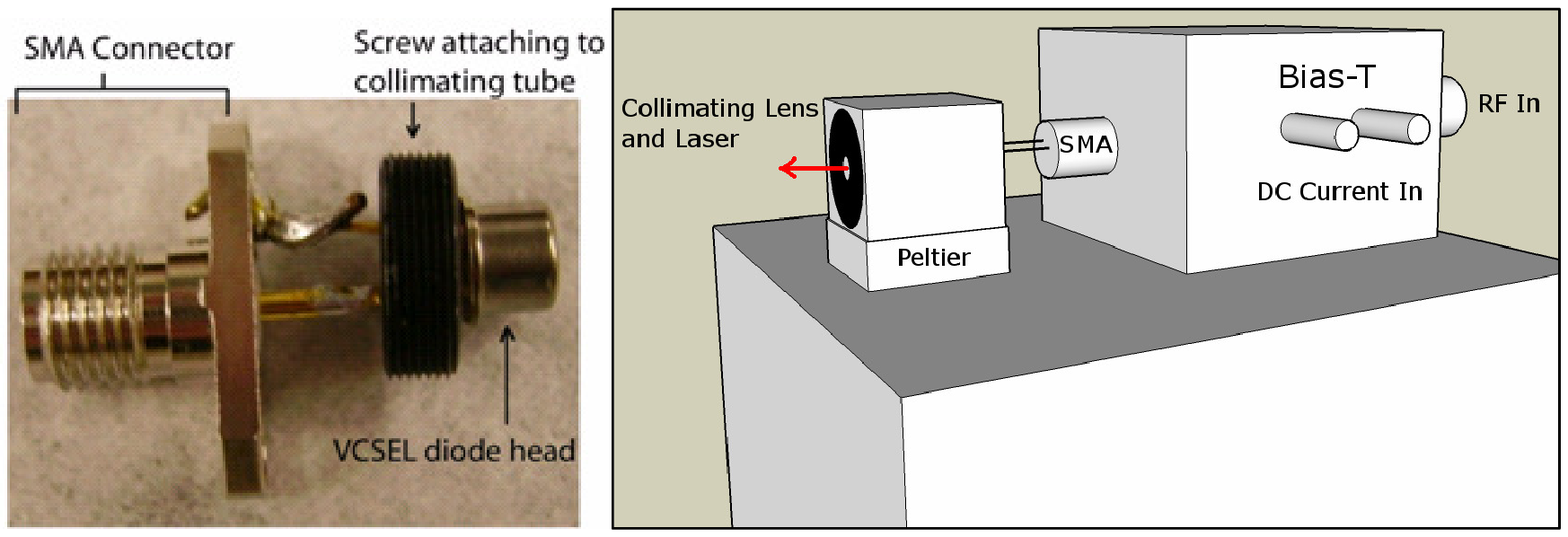}
\caption{\label{vcselassembly.fig} (a) A VCSEL attached to an SMA connector. (b) The schematics of the laser
head assembly. }
\end{figure*}

The level structure of the ${}^{87}$Rb $\mathrm{D}_1$ line is shown in Fig.~\ref{LambdaSystem.fig}(c). Both the
ground ($5S_{1/2}$) and the excited ($5P_{1/2}$) states are split due to the coupling of the electron angular
momenta and the nuclear spin, and each state is labeled with the value of the total angular momentum $F$. The
observation of coherent population trapping requires two laser fields to couple both ground states $F=1$ and
$F=2$ with the same excited state ($F'=2$ in our experiments). However, it is impossible to use two independent
diode lasers due to their relatively large intrinsic frequency noise. On one hand, a dark state
(\ref{darkstate}) exists only if the differential frequency of two laser fields matches the ground state
splitting with good precision (typically better than a few kHz). On the other hand, the electromagnetic field
emitted by a laser is not truly monochromatic, but instead its frequency ``jumps'' randomly in a certain range,
called the laser linewidth. Typical linewidths of commonly used diode lasers are from 10--100\,MHz for a
free-running diode laser to several hundred MHz for a vertical cavity surface emitting diode laser (VCSEL). Even
more sophisticated commercial external-cavity diode lasers have linewidths of the order of 1\,MHz. As a result,
the two-photon detuning of two independent lasers fluctuates in the range determined by the laser linewidths,
and no narrow resonances can be observed.

To avoid this problem we obtain several electromagnetic fields from a single laser by modulating its phase
$\varphi (t) = \epsilon \sin (\omega _m t)$. Such phase modulation is equivalent to producing a frequency comb
with the frequency separation between the ``teeth'' equal to the modulation frequency $\omega_m$, and the
amplitude of each component determined by the phase modulation amplitude $\epsilon$:\cite{mathbook}
\begin{subequations}
\label{phasemod}
\begin{align}
E(z,t)&=\frac12{\mathcal{E} e^{ikz-i\omega t + i \varphi (t)} + c.c.} \\
&=\frac12{ \mathcal{E} \sum ^\infty _{n=0} J_n(\epsilon) e^{ikx - i(\omega - n\omega_m)t} + c.c.}
\end{align}
\end{subequations}
In our experiments we use phase modulation frequency $\omega_m$ to be close to the hyperfine splitting frequency
in ${}^{87}$Rb $\Delta_{\mathrm{hfs}}=6.835$\,GHz. We then use two of the resulting frequency comb components to
form a $\Lambda$ system: the zeroth (at the carrier frequency $\omega$) and one of the first (at frequency
$\omega \pm \omega_m$) modulation sidebands. It is also possible to phase modulate the laser field at half of
the hyperfine splitting frequency, and use two first modulation sidebands to achieve coherent population
trapping.\cite{vanier05apb} A special type of diode laser (a VCSEL)\cite{wynandsvcsel} allows efficient phase
modulation at the desired high microwave frequency by directly modulating its driving current. For this type of
laser an active region (where the lasing occurs) is smaller than in conventional edge-emitting diode lasers,
providing two main advantages: fast response (up to 10\,GHz modulation was achieved for our sample), and very
low power consumption. Both of these properties make a VCSEL the laser of choice for miniature atomic clock
applications.\cite{vanier05apb}

In the following we give a detailed description of the experimental apparatus in our
laboratory.\cite{NathanThesis} An approximate budget for the experiment is provided in Table I.

\begin{table*}
\caption{We list the sources for the important components of the proposed experiments and current prices. The
list does not include raw materials (such as aluminum or acrylic sheets or prototyping circuit boards).}
\begin{tabular*}{\textwidth}{p{6.2cm} p{8cm} p{2cm} p{2cm}}
 \hline
\emph{ Component} & \emph{Source} & \emph{Price} & \emph{Quantity} \\
\textbf{Laser assembly} & & & \\
VCSEL laser @795\,nm& ULM795-01-TN-S46FOP, Laser components & \$400 & 1 \\
for different wavelengths & VCSEL-780 or -850 Thorlabs & \$20 & \\
Collimating tube & LDM-3756, Optima Precision & \$75 & 1 \\
Bias-T & ZX85-12G+, Mini-Circuits& \$100 & 1 \\
Temperature controller & WTC3293, Wavelength Electronics & \$500 & 1\\
TEC element & 03111-5L31-03CP, Custom Thermoelectric & \$20 & 1\\
Optical isolator (optional) & IO-D-780-VLP, Thorlabs & \$460 & 1 \\
\textbf{Rubidium cell enclosure} &&& \\
Rb cell: isotopically \\enriched/natural abundance & Triad Technology & \$625/\$310 & $1^*$ \\
Bifiler heating wire & 2HN063B-13 Ari Industries & \$4/ft & $10'$ \\
\textit{Magnetic shielding} &&& \\
Custom 3 layer design & Magnetic Shield Corporation & \$1300 & 1 \\
alternative: lab kit & & \$100 & \\
\textbf{Microwave equipment} &&& \\
Tunable Mini-YIG Oscillator & Stellex, purchased used on eBay & \$50 & 1 \\
Linear PLL chip evaluation board & LMX2487EVAL National Semiconductor& \$176 & 1 \\
Voltage controlled oscillator & CRO6835ZZ~Communications & \$75 & 1 \\
10 MHz reference oscillator & 501-04609A Streamline & \$305 & 1 \\
Output rf amplifier Amplifier & ZJL-7G Mini-Circuits & \$100 & 1 \\
Various attenuators & VAT-x Mini-Circuits VAT-x & \$16/each & 3--4 \\
Directional coupler & 780-20-6.000 MECA Electronics & \$165 & 1\\
\multicolumn{4}{l}{\textbf{Various optics and opto-mechanics}} \\
Mirrors &Thorlabs (various options) & \$15--\$50 & 2--3 \\
Quarter wave-plate & WPMQ05M-780 Thorlabs & \$230 & $1^*$ \\
Photodetector & Thorlabs (various options)& \$20--\$100 & $1^*$ \\
Optical mounts \& holders & Thorlabs & $\approx \$300$ & \\
\textbf{General lab equipment} &&& \\
\multicolumn{4}{p{16cm}}{Oscilloscope, function generator,
lock-in amplifier, frequency counter, constant $\pm 15$\,V and $\pm 5$\,V and variable power supplies, multimeters.} \\
\hline
\multicolumn{4}{p{16cm}}{${}^*$ Additional items are needed if DAVLL laser lock is used.} \\

\end{tabular*}
\end{table*}

\textit{Laser assembly}. Figure~\ref{vcselassembly.fig} shows the homemade laser head assembly used in
our experiment. A VCSEL (ULM795-01-TN-S46FOP from U-L-M Photonics) is placed inside a
collimating tube (model LDM 3756 from Optima Precision) and a copper holder. The resulting assembly
is attached to a peltier thermoelectric cooler connected to a temperature controller (model
WTC3293-14001-A from Wavelength Electronics) to actively stabilize the diode temperature with
precision better than $0.1^\circ$C. The basis for the laser system is the heat sink, which doubles as
the holding block for the system. The sensitivity of the laser frequency to temperature
($0.06$\,nm/$^\circ$C for our laser) introduces a way to tune the laser to the atomic resonance
frequency, but also puts stringent requirements on the diode's temperature stability. To prevent
temperature fluctuations due to air currents, the laser mount is enclosed in a small aluminum box.

\begin{figure*}[h!]
\centering
\includegraphics[width=0.9\textwidth]{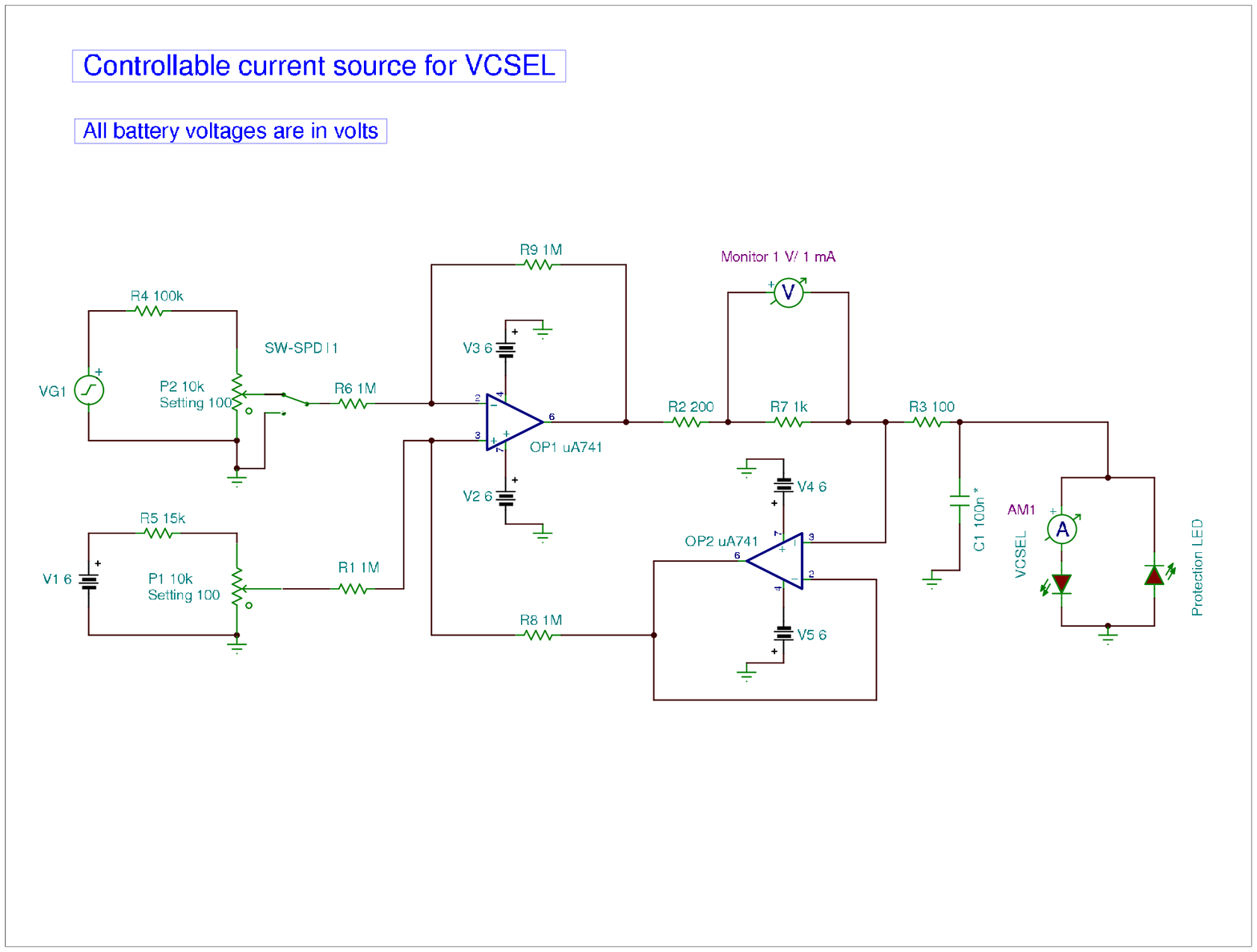}
\caption{\label{VCSEL_Constant_Current_Source.fig}Constant current source for a VCSEL laser. Maximum available
current is 2\,mA.}
\end{figure*}

The laser diode pins are soldered to a standard SMA connector and plugged into a Bias-T (model ZFBT-6GW from
Mini-Circuits), which combines a high-frequency modulation signal and a constant current, required to drive the
VCSEL. A typical driving current required to power a VCSEL is very small (the maximum allowed current is 3\,mA
in our sample). To extend the lifetime of the diode, we operated it at 1--1.5\,mA, resulting in an output
optical laser power of $\approx 0.5$\,mW, collimated to a $1$\,mm beam. Any variations in laser current also
cause fluctuations in the output laser frequency ($\approx 0.9$\,nm/mA). We have designed a simple,
battery-operated, low-noise current supply optimized to drive a VCSEL
(Figure~\ref{VCSEL_Constant_Current_Source.fig} provides the schematics.) For the coherent population trapping
and clock measurements, the output laser frequency is also actively locked to the atomic transition using a
DAVLL (Dichroic Atomic Vapor Laser Locking) technique.\cite{davllwieman,DAVLLbudker}

\textit{Microwave modulation}. In addition to a constant driving current, a high-frequency signal has
to be mixed
in to phase-modulate the output of the laser and produce the frequency comb given by Eq.~(\ref{phasemod}).
In our experiments we tested several microwave oscillators. Our initial tests used
a commercial frequency synthesizer (Agilent E8257D) available in our laboratory. For all
later work it was replaced with less expensive options: a current-tunable crystal oscillator and an
electronic phase-locked loop (PLL) with external 10\,MHz reference.

\begin{figure*}[h!]
\centering
\includegraphics[width=1.0\textwidth]{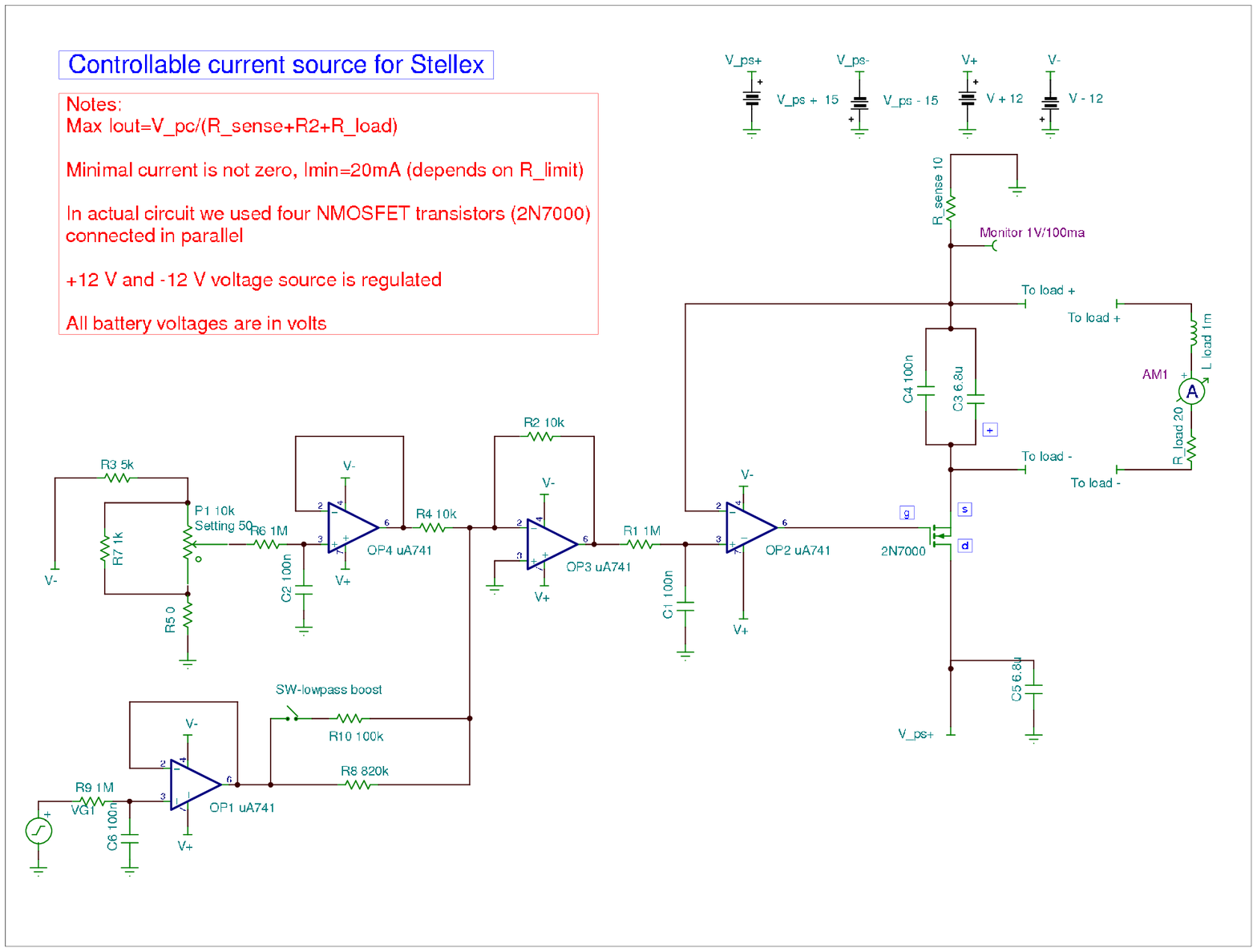}
\caption{\label{Stellex_Oscillator_Constant_Current_Source.fig}Constant current source for a microwave
oscillator tuning coil. Maximum available current is 150\,mA.}
\end{figure*}
\begin{figure*}[h!]
\centering
\includegraphics[width=1.0\textwidth]{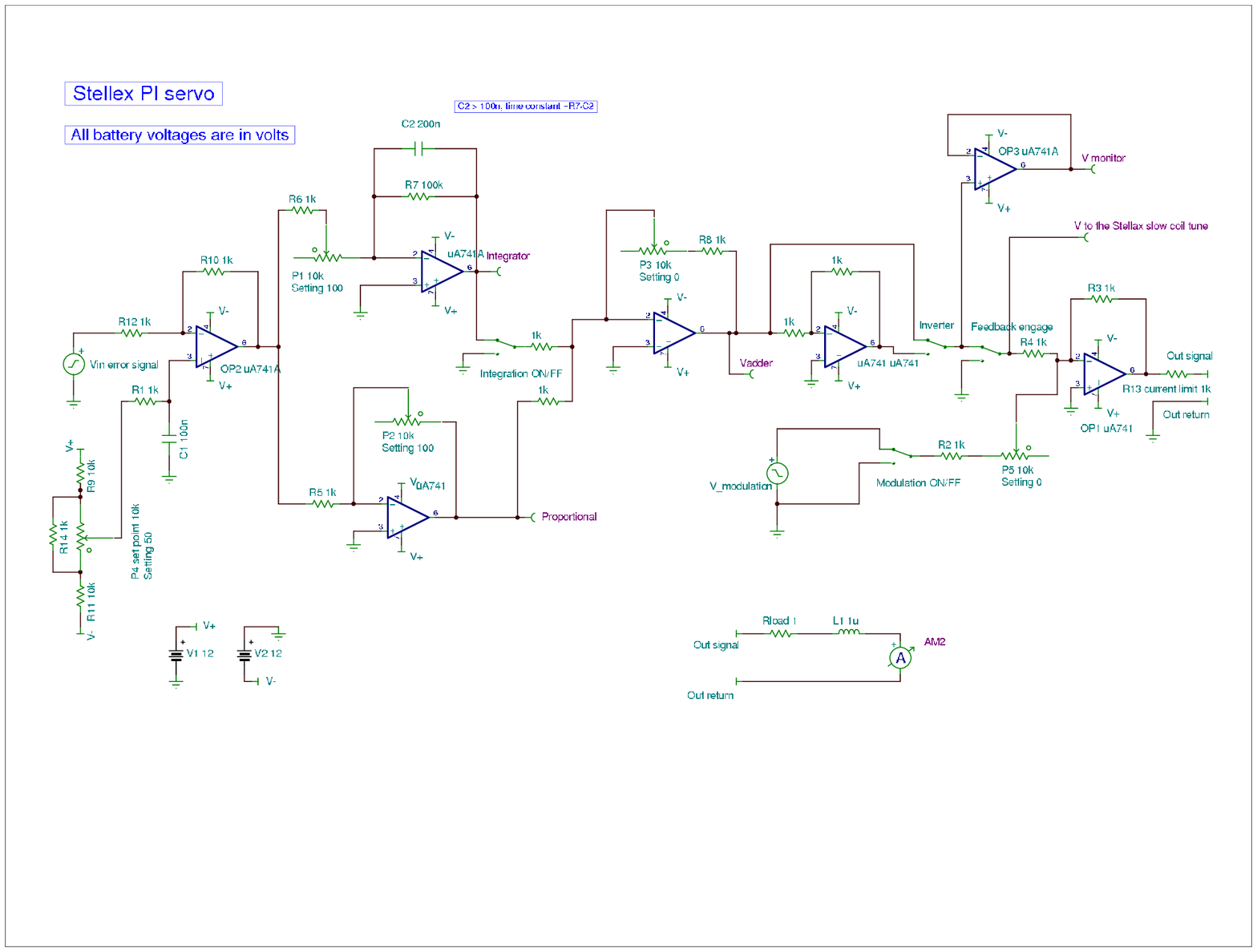}
\caption{\label{PI_Lock_Servo_for_Stellex_Oscillator.fig}A proportional–integral (PI) servo controller for the
oscillator frequency lock.}
\end{figure*}

The most affordable rf source was a Stellex Mini-YIG oscillator purchased from a surplus electronic seller. This
oscillator outputs 15\,dBm of power at a frequency between 5.95\,GHz and 7.15\,GHz. The output frequency of the
oscillator can be tuned by changing the current flowing through two internal magnetic coils: a main coil,
designed for coarse tuning of the oscillator frequency (5\,MHz/mA) at a slow rate up to 10\,kHz, and an
auxiliary fast tuning coil that allows finer frequency control (150\,kHz/mA) with the 400\,kHz tuning bandwidth.
As for the VCSEL driver, a current source used for tuning must have minimal internal noise (better than many
commercial laser current drivers), because any modulation frequency fluctuations are immediately converted into
fluctuations of the two-photon detuning. In our case the tuning current required to reach the designed
oscillator frequency ($6.835$\,GHz) is greater than 100\,mA. To simplify the construction of a tunable low-noise
high-current source, we used a two-stage design. The oscillator frequency is coarsely set to the required value
by manually adjusting a cw current source driving the main tuning coil to approximately 140\,mA, and an
additional tunable low-current source plugged into the auxiliary fast tuning coil is responsible for fine
frequency adjustments in the narrow range of about 1\,MHz around the set value. Circuits for the oscillator
current driver and for the active locking servo are shown in
Figs.~\ref{Stellex_Oscillator_Constant_Current_Source.fig} and \ref{PI_Lock_Servo_for_Stellex_Oscillator.fig}.
The tuning source allows tuning the oscillator frequency to a particular microwave frequency or sweeping the
modulation frequency to observe changes in the optical transmission as a function of the two-photon detuning.
Although all the experiments we discuss can be performed using the Stellex oscillator, it is sometimes not very
convenient to use because of its large internal frequency jitter at the level of several tens of kilohertz.
Also, without active frequency locking, its frequency drifts several hundreds of kilohertz during an hour,
probably because of its poor temperature stability.

The other microwave source we used requires more initial investments, but provides much greater precision and
stability in the output frequency. It is based on the Linear PLL chip (LMX2487) evaluation board (LMX2487EVAL),
which sets a user provided voltage controlled oscillator at an arbitrary frequency near 6.835\,GHz with
sub-Hertz resolution (we used CRO6835Z). The output signal is phased-locked to a 10\,MHz reference. This feature
can be used, for example, to lock the rf output to a stable frequency reference (such as a benchtop Rb frequency
standard FS725 from Stanford Research Systems). Also, we can slightly vary the frequency of a voltage-controlled
10\,MHz oscillator to achieve sweeping capability in the rf output frequency. We used a Streamline oscillator
(501-04609A) with a voltage tuning response of $\pm5$\,Hz, corresponding to a variation in the microwave
frequency of $\pm3$\,kHz. This tuning is sufficient to lock the rf output to the coherent population trapping
resonance and study its stability in an atomic clock arrangement.

\begin{figure}[h!]
\centering
\includegraphics[width=1.0\columnwidth]{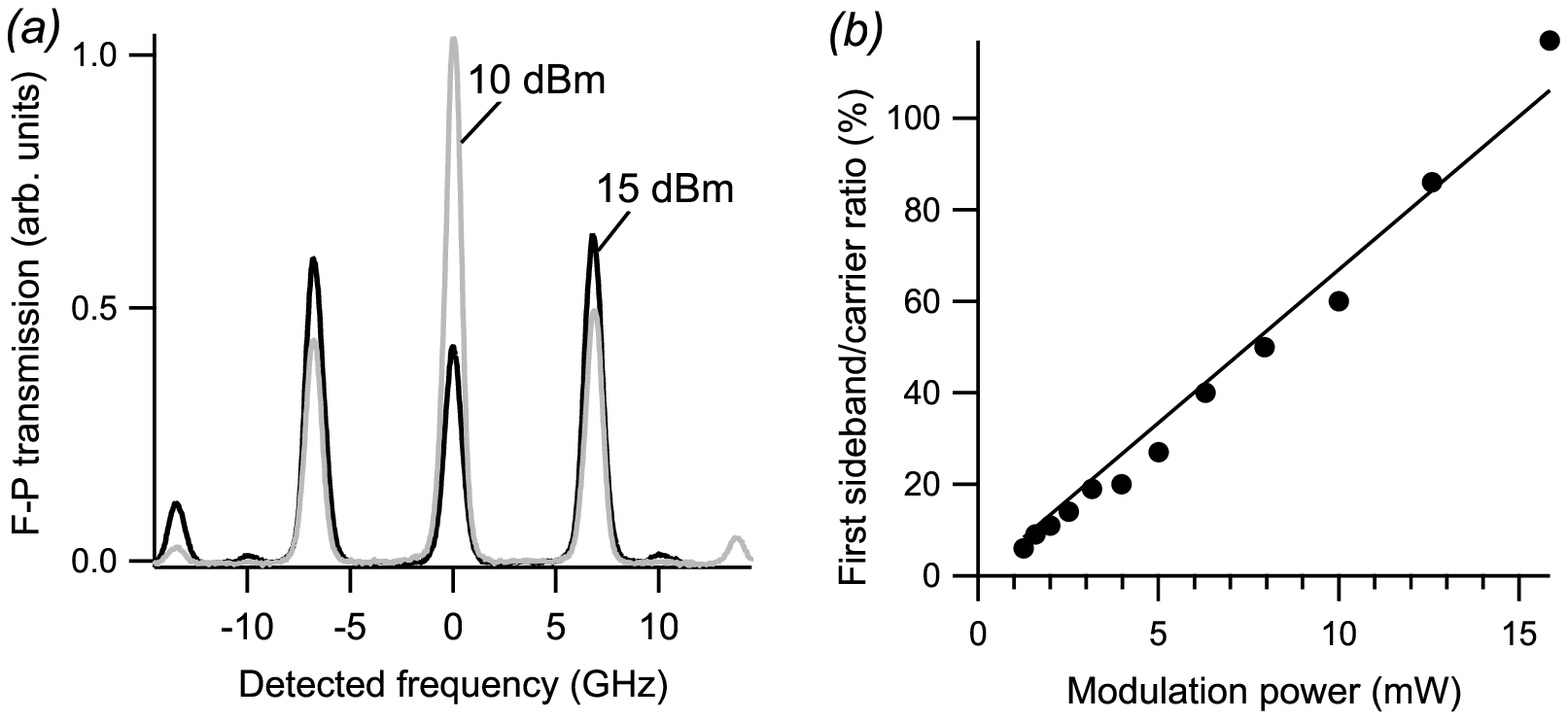}
\caption[vcselsetup]{(a) Frequency spectrum of the VCSEL modulated with $15$\,dBm and $10$\,dBm of rf power. (b)
The dependence of the first sideband/carrier intensity ratio as a function of the rf power.}
\label{rfmodulation.fig}
\end{figure}

The microwave output of either oscillator is connected to the laser through the bias-T. The power of the
oscillator determines the strength of the modulation. Figure~\ref{rfmodulation.fig}(a) shows the modulation comb
for two values of rf power recorded by passing the laser output through a hand made Fabry-Perot cavity with
40\,GHz free spectral range. Figure~\ref{rfmodulation.fig}(b) shows the ratio between the first and the zeroth
(carrier) modulation sidebands, which grows proportionally to the microwave power sent to the VCSEL. Due to the
high efficiency of the high-frequency current modulation, we were able to transfer a large fraction of the
optical power in the first modulation sideband, achieving the first sideband/carrier ratio $>1$ for moderate
modulation power $\ge 15$\,mW (12\,dBm).

\begin{figure}[h!]
\centering
\includegraphics[width=1.0\columnwidth]{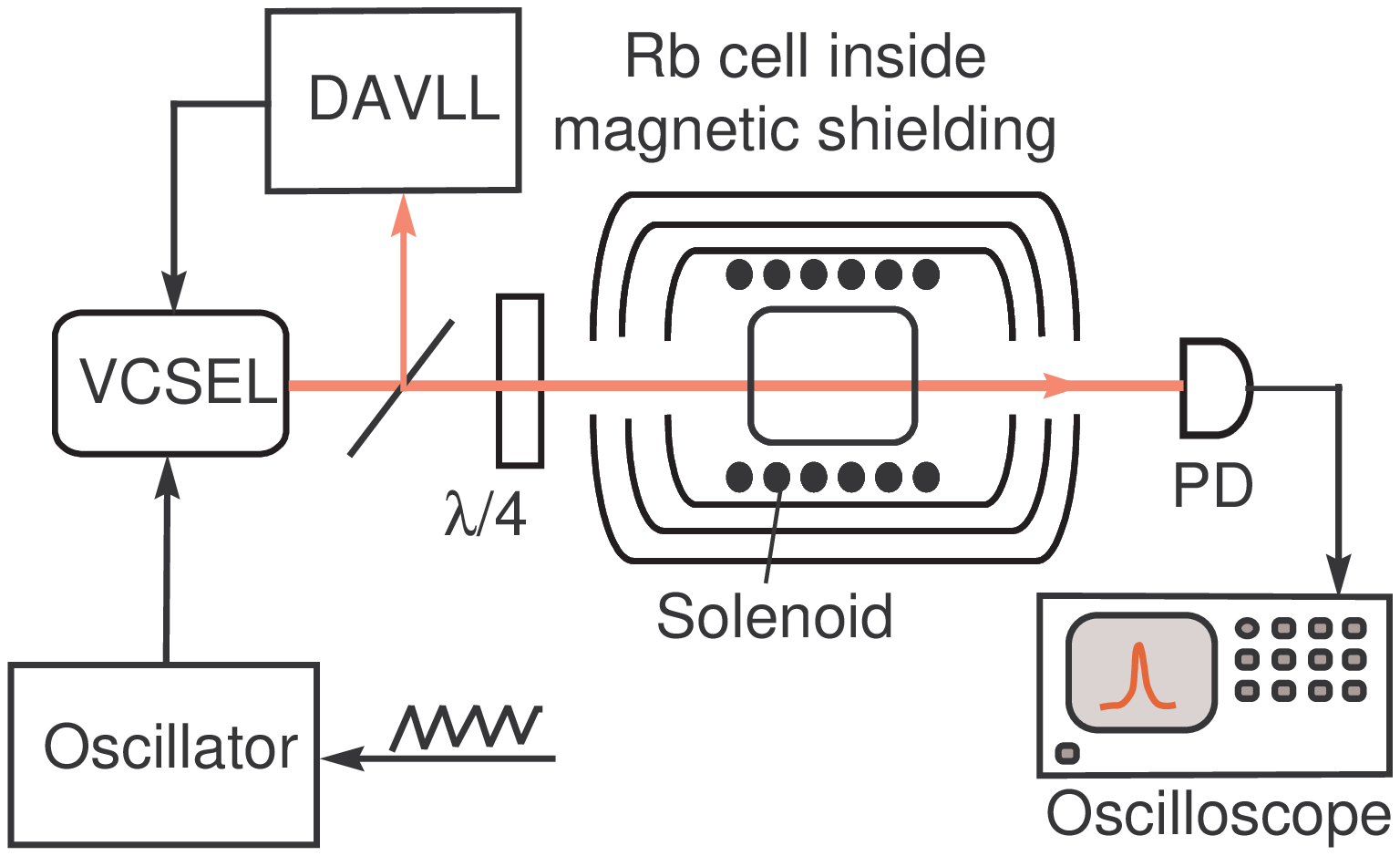}
\caption{\label{setup_EIT.fig}Experimental setup for electromagnetically induced transparency observation. The
radiation of a diode laser (VCSEL) passes through a quater-wave plate ($\lambda/4$) and the shielded Rb cell.
Any changes in its intensity are measured using a photodiode (PD) and a digital oscilloscope. The driving
current of the laser is directly modulated using a tunable microwave oscillator. An additional laser lock
(DAVLL) allows the laser frequency to be maintained at the desired optical transition.}
\end{figure}

\textit{Rubidium cell enclosure}. The remaining important component of the experimental setup is a Rubidium
cell. In our experiment we used a standard cylindrical glass cell ($25$\,mm diameter, $75$\,mm length)
containing isotopically pure $^{87}$Rb (TT-RB87/Ne-75-Q from Triad Technology). It is easy to estimate that at
room temperature, Rb atoms inside the cell move with average speed of $400$\,m/s, and thus it takes only
$2\,\mu$s to cross the $1$\,mm-wide laser beam.\cite{timeofflightnote} After that time an atom most likely
collides with the cell wall, and its quantum state created during the interaction with the laser fields is
destroyed. To amend the situation, our cell also contains $5$\,Torr of Ne buffer gas. Collisions with buffer gas
atoms have a very weak effect on the atomic quantum state, but rapidly change the velocity of Rb atoms,
restricting their motion to slow diffusion. As a result, Rb atoms spend a much longer time inside the
interaction region, effectively increasing their dark state lifetime. For example, for Rb atoms in Ne buffer
gas, the diffusion time of an atom through the laser beam diameter $a$ can be calculated by solving the
diffusion equation. The solution is given by\cite{arimondoPRA96}
\begin{equation} \label{gammadiffusion}
\tau_{\mathrm{diff}}=\frac{a^2}{1.15\,[\mathrm{cm^2/s}]}\frac{P_{{\rm Ne}}}{P_\mathrm{atm}},
\end{equation}
where $P_{\rm Ne}$ is the pressure of the Ne buffer gas inside the cell, and $P_\mathrm{atm}$ is the
atmospheric pressure. For example, for the $1$\,mm beam used in the experiment, $5$\,Torr of Ne buffer
gas extends the interaction time from $2.5\,\mu$s to $60\,\mu$s.
The exact total amount and composition of
a buffer gas is not critical. Any cell with 5--50\,Torr of any inert gas (Ne, He, Ar, Xe) or some
simple diatomic molecules (N$_2$, CH$_4$) is suitable for the experiments described in the next
sections.

Any longitudinal magnetic field splits the magnetic sublevels of all hyperfine states due to the
Zeeman effect, and thereby changes their two-photon resonance frequencies. To avoid stray fields from
the laboratory environment, the Rubidium cell is placed inside a set of three cylindrical magnetic
shields (from Magnetic Shield Corporation) to suppress any external magnetic field by a factor of
$10^3$--$10^4$. Much simpler single-layer shielding may be sufficient to observe coherent
population trapping resonances
of a few tens of kHz wide. It might also be useful to install a solenoid inside the shielding to
change the longitudinal magnetic field and study its effect on coherent
population trapping.

The number of Rb atoms interacting with light (that is, the Rb vapor density inside the cell) is
determined by the saturated vapor pressure with the solid or liquid Rb metal droplet placed inside
the cell, and can be controlled by changing the cell's temperature.\cite{steck} In our experiment
we control the temperature of the cell by passing constant electrical current through a resistive
heater wrapped around the inner magnetic shield. To avoid any magnetic field due to the heater
current, we used a bifiler chromium wire as a heating element. Alternatively, any twisted loop of
wire can be used, because the current will run in opposite directions along each point of the wire.
We found that the coherent
population trapping resonances have the highest contrast at some optimal range of temperatures
(35--50$^\circ$C for our experiment). If the temperature is too low, almost all the light gets through
regardless of coherent
population trapping conditions. If the temperature is too high, it is difficult to see coherent
population trapping resonances due
to strong absorption.

\section{Observation and characterization of coherent
population trapping resonances}

\begin{figure}[h!]
\centering
\includegraphics[width=1.0\columnwidth]{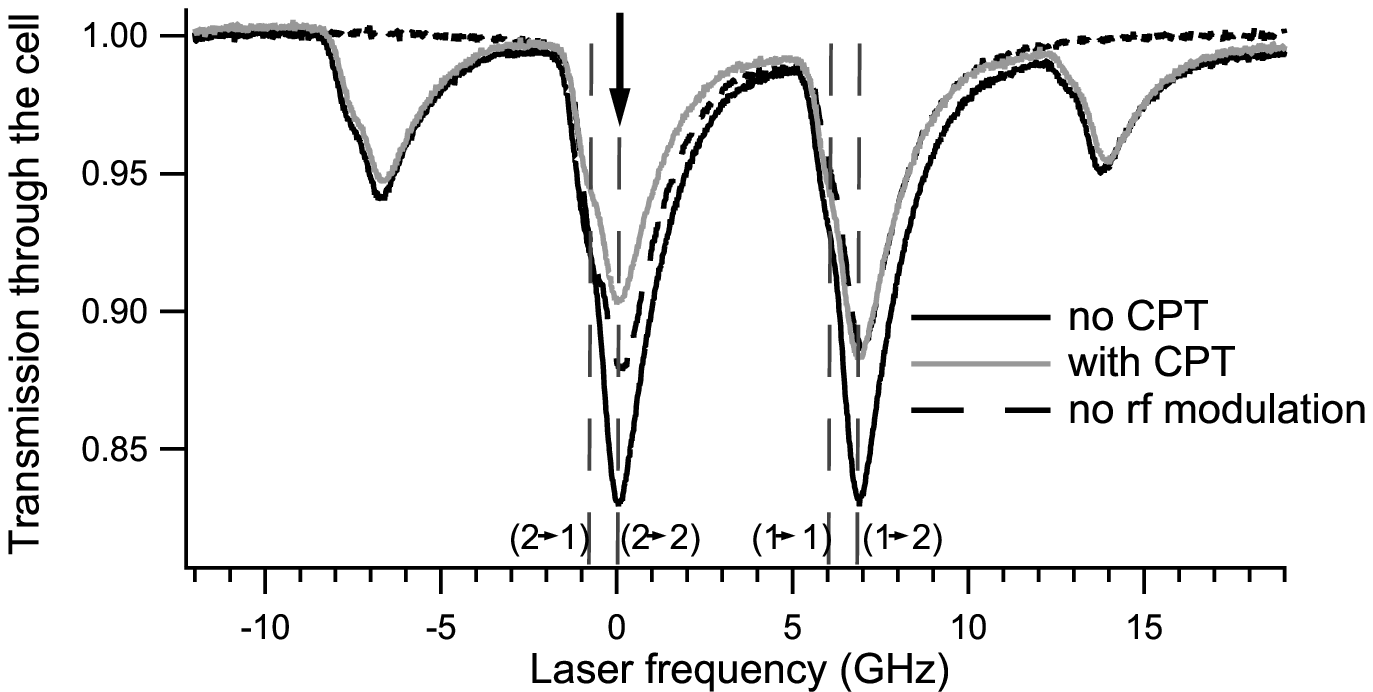}
\caption[cpt]{Dependence of total laser transmission through the cell without rf modulation (dashed line) and
with the rf modulation frequency tuned to the coherent population trapping resonance conditions (solid grey
line) and $2$\,MHz away from the coherent population trapping resonance (solid black line). The sideband/carrier
ratio is 20\%, and the optical laser power is $150\,\mu$W. The positions of all four optical transitions for the
carrier field are labeled and marked with vertical lines. Vertical black arrow indicates the laser frequency
used in all following measurements. \label{abs_cpt.fig}}
\end{figure}

Our experimental apparatus enables a range of experiments on the coherent properties of
atoms and their applications. The first and basic one is the observation of coherent
population trapping resonances by
measuring the transmission of the laser light through the atomic cell while scanning the laser
modulation frequency over the two-photon resonance. The first step in this process is tuning the
laser to the right optical frequency. Figure~\ref{abs_cpt.fig} shows the transmission of the laser
light when its frequency is swept across all four optical transitions of the ${}^{87}$Rb $\mathrm{D}_1$
line [the transitions are shown in Fig.~\ref{LambdaSystem.fig}(c)]. The spectral width of each
absorption line is determined by the decoherence rate of the optical transitions of approximately
$600$\,MHz FWHM (dominated by the Doppler broadening). As a result, the two transitions from the same
ground state to each of the excited states are not completely resolved, because their relatively small
hyperfine splitting ($\approx 800$\,MHz) is comparable to the width of each individual transition. The
transitions from each of two ground states are clearly separated due to significantly larger
ground-state hyperfine splitting ($\Delta_{\mathrm{hfs}}=6.835$\,GHz).

The output optical frequency of the laser is sensitive to the rf modulation power, and the precise tuning
should be performed with modulation turned on. The tuning process is more convenient if the optical
power in each of the first modulation sidebands is $\approx 20$--60\% of the carrier field, so it is
easy to distinguish the carrier and sideband absorption. Then the rf oscillator frequency should be
appropriately tuned -- first with the coarse tuning to a value close to the hyperfine splitting, and
then with fine adjustments such that a clear increase in light transmission is observed near Rb
resonances, as illustrated by the difference in the solid lines in Fig.~\ref{abs_cpt.fig}. To achieve
the maximum contrast of coherent
population trapping resonances with circularly polarized light, the carrier (that is, the
unperturbed laser frequency) should be tuned to the $F=2 \rightarrow F^\prime = 2$ transition. In
this case, the high-frequency modulation sideband is resonant with the $F=1 \rightarrow F^\prime =
2$. This arrangement is indicated by an arrow in Fig.~\ref{abs_cpt.fig}.

\begin{figure}[h!]
\centering
\includegraphics[width=1.0\columnwidth]{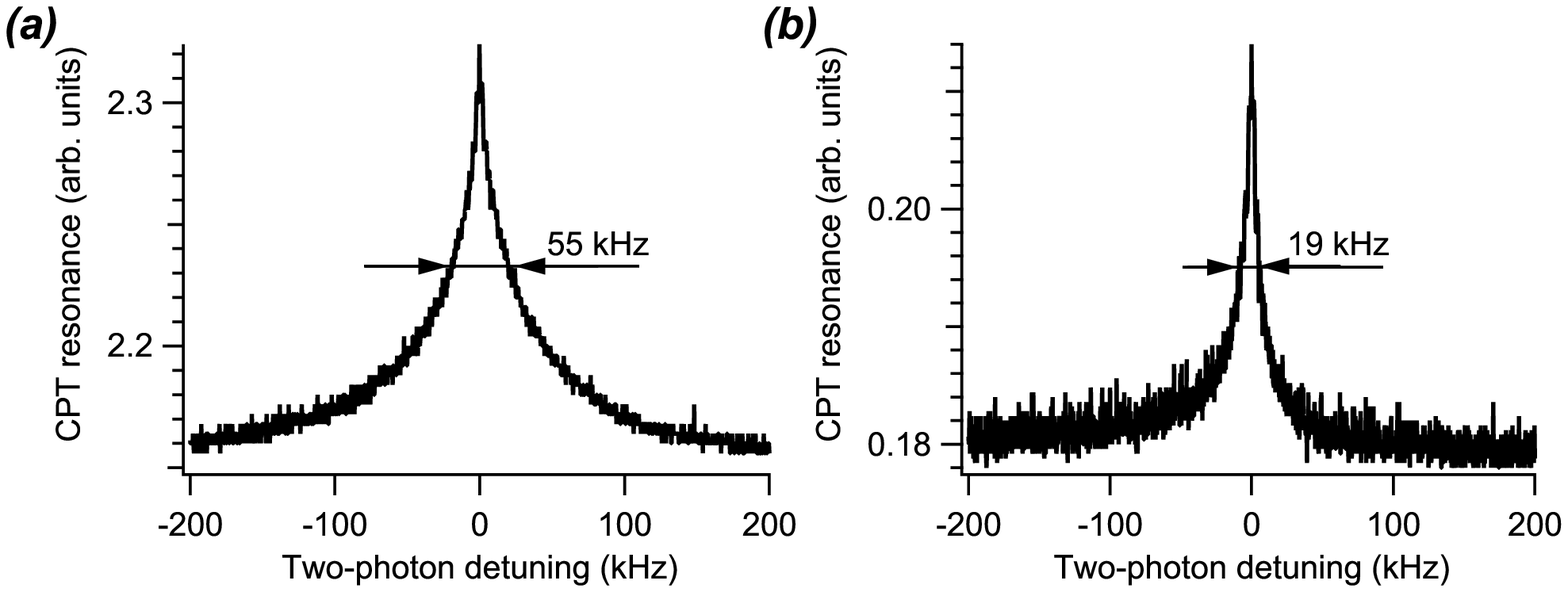}
\caption{Coherent population trapping transmission resonances observed by sweeping the rf modulation frequency
in a $400$\,kHz range around 6.834685\,GHz for two values of the optical laser power: (a) $132\,\mu$W and (b)
$32\,\mu$W.} \label{cpt.fig}
\end{figure}

Once the laser optical frequency is parked at the right transition, we can directly observe a coherent
population trapping
resonance by slowly scanning the modulation frequency near the two-photon transition.
Figure~\ref{cpt.fig} shows examples of narrow (a few kHz) peaks in the transmitted light power after
the Rb cell when the coherent
population trapping condition is fulfilled. Because all the laser light is detected, the coherent
population trapping
transmission peak is observed on top of an often large background. This background transmission
consists of the off-resonant modulation sidebands, which do not interact with atoms, as well as
residual transmission of the resonant light fields, if the atomic density is not too high. It is also
easy to see that even the maximum of the coherent
population trapping resonance does not reach 100\% transmission, because the
lifetime of the dark state is finite, and there is always a fraction of atoms in the laser beam that
is absorbing light. The exact position of the transmission maximum depends on many parameters, such
as the power of the light (``light shift''), and the amount and the temperature of the buffer gas
(``pressure shift''). For example, in the cell we used in the experiments (filled with $5$\,Torr of
Ne) the position of the two-photon resonance at low light level was measured to be $6.834685$\,GHz.

After initial detection of coherent population trapping resonances, it may be useful to systematically study
important properties such as the dependence on the laser field strength. For example, Eq.~(\ref{gammawidth})
predicts the broadening of coherent population trapping resonances at higher laser powers. It is easy to verify
this broadening experimentally by measuring the resonance width while reducing the light power interacting with
atoms by placing (for example) a few neutral density filters in front of the cell. The example of the resonance
narrowing is shown in Fig.~\ref{cpt.fig}. Note that the resonance linewidth may not follow
Eq.~(\ref{gammawidth}) exactly, especially for moderate buffer gas pressure ($\le 10$\,Torr), because the
dynamics of atomic diffusion becomes important. In particular, if atoms are allowed to diffuse out of the laser
beam and then return without losing their quantum state,\cite{xiaoPRL06} the width of the coherent population
trapping resonances may become much narrower than expected from the simple diffusion picture provided by
Eq.~(\ref{gammadiffusion}).

We can also observe that the coherent
population trapping resonance lineshape becomes asymmetric when the optical laser
frequency is detuned from the exact optical transition frequency.\cite{mikhailovPRA04} This effect can
be used to fine tune the laser frequency to the resonance position, which corresponds to the most
symmetric coherent
population trapping resonance. At the same time, students may find it interesting to observe this
systematic lineshape change, accompanied by the reduction of the resonance amplitude and growth in
background transmission as fewer and fewer atoms interact with the laser fields.

\section{Effect of the Zeeman structure on coherent
population trapping resonances; atomic magnetometer}

So far we have ignored the fact that each hyperfine sublevel $F$ consists of $2F+1$ magnetic (Zeeman)
sublevels characterized by $m_F$. These sublevels are degenerate in zero magnetic field, but if a magnetic field is
applied along the light propagation direction (for example, by running the current through the
solenoid mounted inside the magnetic shielding), the sublevels shift by an amount proportional to the applied
magnetic field $B$:
\begin{equation}
\delta_{m_F}=m_Fg_FB,
\end{equation}
where $m_F$ is the magnetic quantum number for each sublevel, and $g_F$ is the gyromagnetic ratio. For Rb levels
the gyromagnetic ratios are very small ($g_{F=1}=0.7$M\,Hz/G and $g_{F=2}=-0.7$\,MHz/G),\cite{steck} and a high
magnetic field is required to shift the levels far enough to resolve individual optical absorption resonances
from different Zeeman sublevels. In contrast, coherent population trapping resonances are much more sensitive to
level shifts.

\begin{figure}[h!]
\centering
\includegraphics[width=1.0\columnwidth]{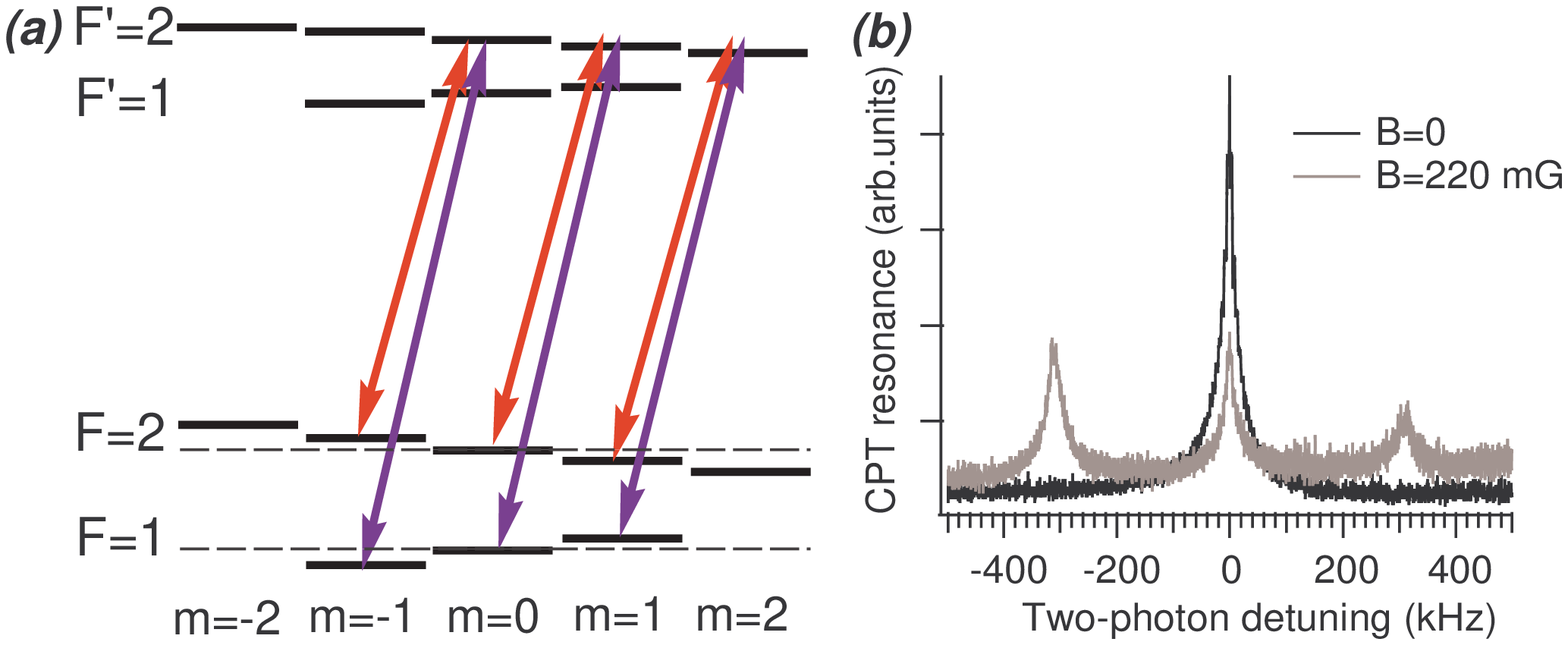}
\caption{\label{Zeeman_cpt.fig} (a) Level diagram for $\mathrm{D}_1$ line of ${}^{87}$Rb showing the Zeeman
sublevels. (b) Coherent population trapping resonance at zero magnetic field and for $B=220$ mG.}
\end{figure}

Let's look more carefully at what happens to a coherent population trapping resonance in the presence of a
magnetic field. Figure~\ref{Zeeman_cpt.fig}(a) shows the relevant Rb energy levels taking into account their
magnetic structure. In this case we must consider not one, but three independent $\Lambda$ systems formed by the
carrier and high frequency sideband fields between three pairs of magnetic sublevels $m_F=0,\pm 1$ in the ground
states and magnetic sublevels $m_F= 0$, 1, and 2 (or $m_F= 0$, $-1$, and $-2$ depending on which circular
polarization is used). When these levels are degenerate (no magnetic field) the dark state is formed at each
pair of ground state sublevels at the same rf frequency $\Delta_{\mathrm{hfs}}$. If the oscillator frequency is
swept around this value, only one coherent population trapping peak is observed. However, in the presence of an
applied magnetic field, the magnetic sublevels with different magnetic quantum numbers $m_F$ shift by different
amounts, as shown in Fig.~\ref{Zeeman_cpt.fig}(a). The conditions for the two-photon resonance at
$\Delta_{\mathrm{hfs}}$ are obeyed only for the non-shifted $m_F=0$ pair, and for $m_F= \pm 1$ levels the
resonance now occurs at $\Delta_{\mathrm{hfs}} \pm 2g_FB$. As a result, the original single coherent population
trapping peak splits into three peaks as shown in Fig.~\ref{Zeeman_cpt.fig}(b). The frequency difference between
the peaks is proportional to the applied magnetic field, and can be used as a sensitive magnetometer.

\section{Coherent
population trapping-based atomic clocks}

Before discussing the application of coherent population trapping resonances for atomic clocks, it is helpful to
mention the definition of the second. The second is the duration of 9\,192\,631\,770 periods of the radiation
corresponding to the transition between the two hyperfine levels of the ground state of the cesium-133
atom.\cite{bipm} This definition implies that we have to accurately measure the frequency between two hyperfine
states of Cs, and then count the number of oscillations to determine the duration of one second. For clocks it
does not matter much if another alkali metal atom is used instead of Cs, because the hyperfine splitting of most
of them is known with great precision. Thus, the principle of an atomic clock is simple. A frequency of some rf
oscillator has to be locked to match the frequency difference between two hyperfine states of Cs or Rb (``clock
transition''), and then the oscillation periods of this locked oscillator can be used as tick marks for
measuring time. The international time standard at the National Institute of Standards and Technology, a Cs
fountain clock, operates on this principle by directly probing the clock transition using microwave radiation.
Extreme care is taken in this case to avoid any systematic errors in measurements to ensure relative frequency
stability approaching $10^{-17}$. However, many practical applications could benefit from an atomic clock whose
stability is at the level of $10^{-12}$--$10^{-13}$, so long as such a device is compact, robust, and
power-efficient. In particular, coherent population trapping resonances are very attractive for the development
of miniature atomic clocks, because all optical components can be miniaturized without loss of
performance,\cite{hollberg04APL} and no bulky rf cavity is needed. Because the coherent population trapping
resonance occurs exactly when the frequency difference between two optical fields matches the hyperfine
splitting in Rb (the clock transition), a coherent population trapping-based atomic clock can be constructed by
locking the rf oscillator used to modulate the VCSEL to the maximum transmission through the atomic cell.

\begin{figure}[h!]
\includegraphics[width=1.0\columnwidth]{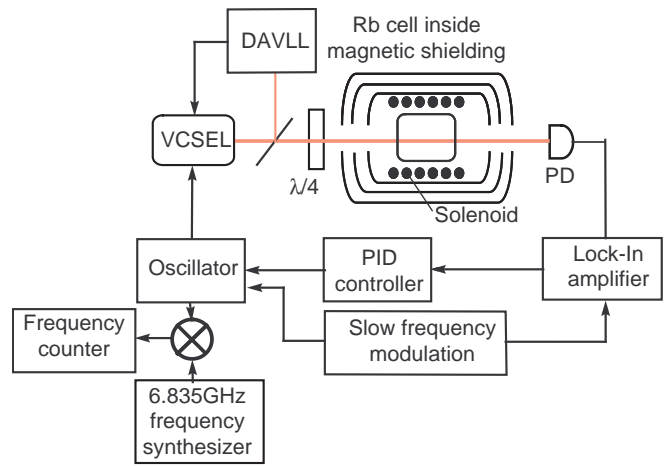}
\caption{\label{setup_clock.fig}Experimental setup for atomic clocks. Most elements are the same as for
Fig.~\ref{setup_EIT.fig}, except that now the output of the photodetector is used in the feedback loop to lock
the microwave oscillator frequency at the peak of a coherent population trapping resonance. A small fraction of
the oscillator output is then mixed with an external reference oscillator to measure the stability of the
resulting atomic clock.}
\end{figure}

To realize such feedback in our experiment, a few additional pieces of equipment were added to the setup, as
shown in Fig.~\ref{setup_clock.fig}. First, a slow dithering modulation (5--10\,kHz) was superimposed on top of
the oscillator rf frequency. When the oscillator frequency is within a coherent population trapping resonance,
this modulation induces a corresponding modulation in the output optical transmission. Phase-sensitive detection
of this signal using a lock-in amplifier transforms a symmetric transmission peak to an anti-symmetric error
signal, which is zero at precisely the maximum of the coherent population trapping resonance. This error signal
is then fed back to the tuning current of the oscillator to correct its frequency and prevent it from drifting
from the clock transition frequency.

To evaluate the performance of the constructed atomic clock we need to measure the frequency
stability of the locked oscillator that outputs the signal at about $6.8$\,GHz. Although it is possible
to measure a few GHz frequency with high enough precision (10 or 11 significant figures), the
required microwave equipment is very expensive. Instead, we split a few percent of the oscillator
output using a directional coupler (780-20-6.000 from MECA Electronics) and mix it with a
stable reference frequency source at a similar frequency.
The resulting beat signal is in the few megazertzs range, and can be accurately measured with the
required precision using a standard frequency counter.

\begin{figure}[h!]
\centering
\includegraphics[width=1.0\columnwidth]{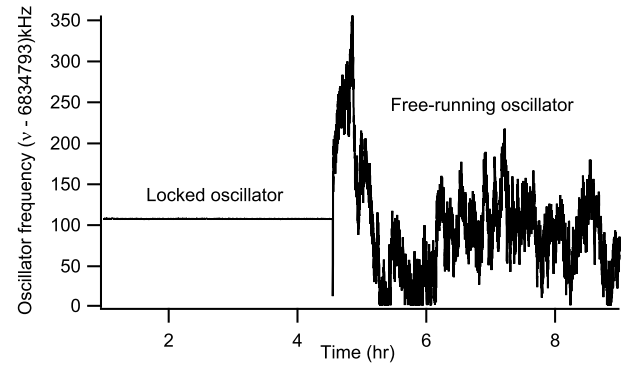}
\caption[lockedvfree]{The oscillator performance when it is locked to coherent population trapping resonance and
when it is free-running.} \label{lockedvfree.fig}
\end{figure}

The difference in performance for the noisy Stellex oscillator when it is free-running or when it is
locked to the coherent
population trapping resonances makes the advantage of a coherent
population trapping-based atomic clock more striking, as
illustrated in Fig.~\ref{lockedvfree.fig}. It is is easy to see a significant difference in
oscillator stability. Although the frequency of the free-running oscillator fluctuates by $\pm 175$\,kHz,
the locked oscillator is stable within $\pm 2$\,Hz.

A common measure of an oscillator's stability is the Allan variance.\cite{allan2} To calculate the
Allan variance we have to divide the entire set of frequency measurements into $n$ sampling periods
of duration $\tau$, and then calculate the average frequency value $\nu_i$ for each interval. The sum of the squared differences between the two consecutive sampling
periods $(\nu_{i+1}-\nu_{i})^2$ is a good quantitative measure of the variation of the average
frequency during the time $\tau$. The definition of the Allan variance is
\begin{equation}
\label{allanvareqn}
\sigma^2 (\tau) = \frac{1}{2(n-1)} \sum_{i=0}^{n-1}(\nu_{i+1} - \nu_i)^2 .
\end{equation}
The Allan variance depends on the averaging time $\tau$ and is usually displayed as a graph. The
lower the Allan variance, the greater the stability of the oscillator. For a very stable but noisy
oscillator the value of the Allan variance monotonically improves with larger $\tau$, because longer
integration time reduces the effect of random noise. However, any long-term systematic drift causes
the Allan variance to grow once the duration of the sampling period becomes comparable to the
characteristic drift time. Typically, the minimum in the Allan variance plot indicates the optimal
averaging time for the given experiment.

\begin{figure}[h!]
\centering
\includegraphics[width=1.0\columnwidth]{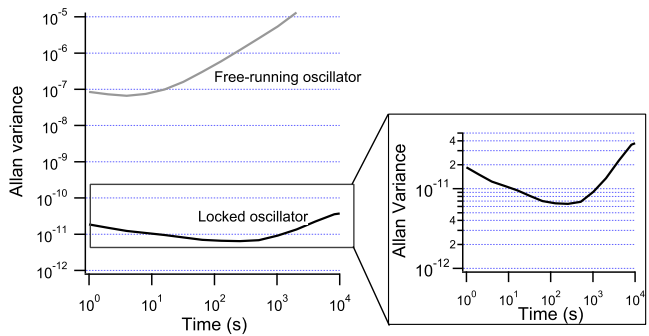}
\caption[allanvar]{Fractional Allan variance $\sigma(\tau)$ with the clock locked (black) and free-running
(grey).} \label{allanvar.fig}
\end{figure}

Figure~\ref{allanvar.fig} shows the Allan variance for the frequencies of the locked and the free-running
oscillator. This plot clearly demonstrates close to four orders of magnitude improvement in the oscillator
performance, from $9 \times 10^{-8}$ over a duration of 10\,s in the free-running regime, to $8 \times 10^{-12}$
over 100\,s in the regime of the coherent population trapping-based atomic clocks. To put this stability into
perspective, a clock would lose 1\,s every 4000 years if we could keep our clock that stable indefinitely. We
also measured how long our clock stays locked to microwave clock transition. Our best attempt yielded a locking
time of 41\,h with a 6\,Hz drift, which was most likely limited by the quality of the DAVLL optical lock.

\section{Conclusions}

We have presented the details of the construction of an affordable and versatile
experimental apparatus for observing coherent effects in an atomic vapor suitable for an
undergraduate laboratory. Assembly and debugging the apparatus is an appropriate task for a
senior research project (the experimental apparatus described here was mainly designed and assembled
by one of the co-authors (NB) during his senior year). Work with
this apparatus will allow students to learn the basics of diode lasers, rf equipment, and atomic
spectroscopy. We also described three experiments that can be realized using this apparatus.
We first described the procedure for observing coherent
population trapping transmission resonance due to manipulations
of the coherent state of atoms. We then took advantage of the extreme sensitivity of the coherent
population trapping
resonance frequency to small shifts in the energy levels of atoms to measure a small magnetic field.
Finally, we locked the rf oscillator frequency to the coherent
population trapping resonance using a feedback mechanism to
create a prototype atomic clock. The same apparatus also can be used for slow and stored light experiments if the amplitude of the microwave radiation can be shaped into
probe pulses.

The ability to manipulate quantum states of an atom using light has led to several important applications, such
as quantum memory and slow light, that are currently one of the most active and interdisciplinary areas of
physics.~\cite{lukin03rmp,boydreview,lukin06opn} Familiarity with the basics of these effects through hand-on
experience would be very beneficial for undergraduate students.

\begin{acknowledgements}
The authors would like to thank Sergey Zibrov for advice on the experimental apparatus design, Chris Carlin for
the help with the laser lock construction, and Nate Phillips and Joe Goldfrank for useful feedback on the
manuscript. This research was supported by Jeffress Research grant J-847, the National Science Foundation
PHY-0758010, and the College of William~\&~Mary.
\end{acknowledgements}

\end{document}